%% file: Ibort.tex
\documentclass[a4paper,11pt,reqno]{amsart}

\usepackage{graphicx}        

\usepackage[english]{babel}

\usepackage{amsmath}

\makeatletter
\newcommand{\definetitlefootnote}[1]{%
  \newcommand\addtitlefootnote{%
    \makebox[0pt][l]{$^{*}$}%
    \footnote{\protect\@titlefootnotetext}
  }%
  \newcommand\@titlefootnotetext{\spaceskip=\z@skip $^{*}$#1}%
}
\makeatother

\begin{document}

\definetitlefootnote{\scshape Dedicate to Alberto Ibort on the occasion of his sixtieth birthday.}
\title[Near-horizon modes]{Near-horizon modes and   self-adjoint extensions of the Schr\"odinger 
operator\addtitlefootnote}

\author{A. P. Balachandran}
\address{Physics Department, Syracuse University, Syracuse, New York 13244-1130, 
USA}
\email{balachandran38@gmail.com}

\author{A. R. de Queiroz}
\address{Instituto de Fisica, Universidade de Brasilia, C.P. 04455, 70919-970  Brasilia, DF, Brazil}
\email{amilcarq@unb.br}

\author{Alberto Saa}
\address{Department of Applied Mathematics, 
 University of Campinas,  13083-859 Campinas, SP, Brazil}
\email{asaa@ime.unicamp.br}
%
%

\maketitle

\begin{abstract}
We investigate the dynamics of   scalar fields in the  near-horizon exterior region of a Schwarzschild black hole. We show that  low-energy modes are typically long-living and might be considered as being confined near  the  black hole horizon. 
Such dynamics are effectively governed   by a Schr\"odinger operator with infinitely many 
self-adjoint extensions parameterized by $U(1)$, a situation closely resembling the  case of an ordinary
free particle moving on a semiaxis.   
Even though these different self-adjoint extensions  lead   to equivalent scattering and thermal processes, a comparison with a simplified model suggests a physical prescription to chose the pertinent 
self-adjoint extensions. However, since  all   extensions are in principle physically equivalent,
they might be considered in equal footing  for statistical analyses of near-horizon modes around black holes.  Analogous results   hold    for any  non-extremal, spherically symmetric,  asymptotically flat black hole.
\end{abstract}

\section{Introduction}

The dynamics of quantum and classical fields in the vicinity of black holes have received considerable
attention recently. Several aspects of the so-called soft photons theorems and the asymptotic symmetries in black hole  spacetimes depend ultimately upon the dynamics and the underlying algebraic structure of test fields in the near-horizon region of black holes. For a recent comprehensive review on these subjects, see, for instance,  \cite{Lecture}. Here, we revisit the case corresponding to the simplest classical configuration of a field in the near-horizon region of a black hole:   a massless Klein-Gordon field $\varphi$ around a
  Schwarzschild   black hole,  which metric in standard coordinates reads
\begin{equation}
\label{SS-metric-1}
 ds^2=-\left( 1 - \frac{2M}{r}\right) dt^2 + \frac{ dr^2}{1 - \frac{2M}{r}} + r^2 d\Omega^2. 
\end{equation}
As we will see,   massive scalar  fields can be easily accommodated in our discussion, without altering our main conclusions.
By exploring the  standard decomposition for the scalar field 
\begin{equation}
\label{decomposition}
\varphi_{\ell m} = \frac{e^{-i\omega t}}{r}u_{\ell m}(r) Y_\ell^m(\theta,\phi)
\end{equation}
and the usual tortoise coordinates
\begin{equation}
r_* = r + 2M\log\left(\frac{r}{2M} -1 \right),
\end{equation}
one has the following effective Schr\"odinger equation for the radial function $u_{\ell m}$
\begin{equation}
\label{Schr0}
\left(-\frac{d^2}{dr_*^2} + V_{\ell}(r) \right) u_{\ell m} = \omega^2 u_{\ell m} ,
\end{equation}
where the effective potential $V_{\ell}(r)$ is given by
\begin{equation}
\label{pot}
V_\ell(r) = \left( 1 - \frac{2M}{r}\right) \left(\frac{\ell(\ell+1)}{r^2} + \frac{2M}{r^3}\right),
\end{equation}
which   well known aspect is depicted in Fig. 1.
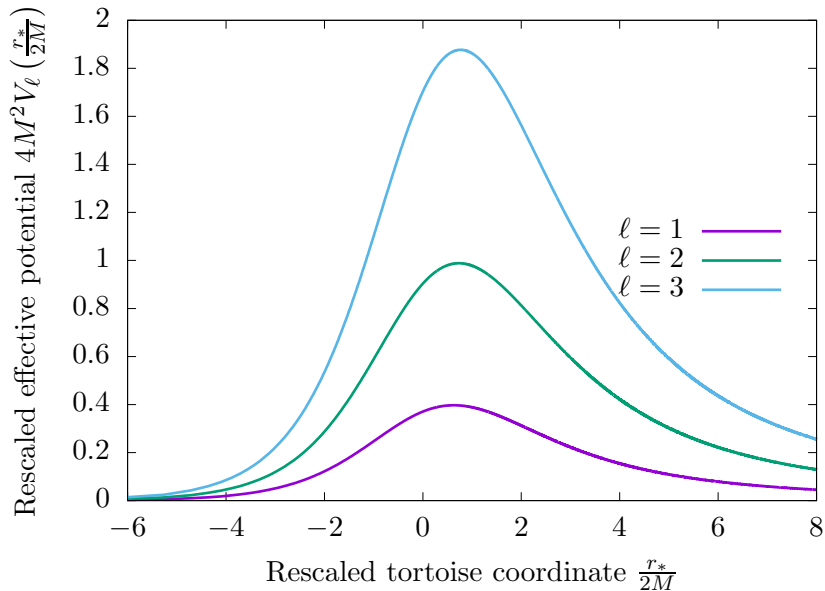
\begin{figure}[t]
\begin{center}
\input{plot.tex}
\end{center}
 \caption{Aspect of the effective potential $V_\ell(r_*)$ given by (\ref{pot}) for some values of $\ell$. The potential decreases exponentially  in the near-horizon region ($r_*\to -\infty$),  see (\ref{exp}),  and as a power law for $r_*\to\infty$ (the asymptotically flat region $r\to\infty$).}
 \label{fig1}
\end{figure}
The tortoise coordinate $r_*$ runs over $(-\infty,\infty)$, with the near-horizon region corresponding to $r\to 2M$ and $r_* \to -\infty$, where the effective potential can be well approximated as
\begin{equation}
\label{exp}
V_\ell(r_*) \approx V^{\rm nh}_\ell(r_*) = \frac{\ell(\ell+1)+1}{4M^2e}\exp\left( \frac{r_*}{2M}  \right).
\end{equation}
For $r\to\infty$, which corresponds to $r_*\to\infty$, the effective potential decreases as a power law. 
For scalar fields with mass $m_\varphi\ne 0$, there will be an extra term $m_\varphi^2$ inside the parenthesis of the second term in (\ref{pot}). It will not alter the effective potential exponential decay in the near-horizon region, nor the power law decay at infinity, although in this case $V_\ell  \to  m_\varphi^2$  for $r\to \infty$. Since the near-horizon potential (\ref{exp}) is not qualitatively 
altered by the mass term, our main conclusions will also hold for the massive case.

The effective Schr\"odinger equation (4) governs all  dynamical  processes involving scalar fields around Schwarzschild black holes. Scattering problems, in particular, involve certain boundary conditions 
at horizon and at infinity.  In these problems, typically, one starts with a incoming  wave from infinity which is scattered by the effective potential barrier    (Fig. \ref{fig1}), leading to a reflected wave towards infinity and a transmitted wave that plunges into the black hole horizon. Such a typical situation corresponds to the following boundary conditions for $u_{\ell m}$
\begin{equation}
u_{\ell m}(r_*) = \left\{ 
\begin{array}{cl}
  A_{\ell m}^{\rm in}(\omega)e^{-i\omega r_*} + 
  A_{\ell m}^{\rm out}(\omega)e^{i\omega r_*}
, & r_*\to \infty ,\\
  B_{\ell m}^{\rm in}(\omega)e^{-i\omega r_*}, & r_*\to -\infty .
\end{array}
\right.
\end{equation}
The (complex) values of $\omega$ such that $A_{\ell m}^{\rm in}(\omega)=0$ are known to correspond to the so-called quasinormal modes, which dominate the asymptotic evolution of non-stationary configurations of the scalar field, see \cite{Review,Review1} for   comprehensive reviews on the subject. 
Here, we are interested in a different field configuration. We will consider processes which originate in the near-horizon region of the black hole and eventually are transmitted to the infinity through the potential barrier. This situation corresponds to the following boundary conditions
\begin{equation}
\label{cond}
u_{\ell m}(r_*) = \left\{ 
\begin{array}{cl}
A^{\rm out}_{\ell m}(\omega)e^{i\omega r_*}
, & r_*\to \infty ,\\
B_{\ell m}^{\rm in}(\omega)e^{-i\omega r_*} +
B_{\ell m}^{\rm out}(\omega)e^{i\omega r_*} , & r_*\to -\infty .
\end{array}
\right.
\end{equation}
We will focus in the lower energy limit, which of course corresponds to small $\omega$, which we assume to be positive. We will discuss the possibility of imaginary $\omega$, which would correspond to  negative eigenvalues $\omega^2$ in the effective Schr\"odinger eigenproblem  (\ref{Schr0}), in the last section. In the low-energy limit, we expect on physical grounds to have some oscillatory behavior in the near-horizon region  and an exponential suppression, due to the effective potential barrier, as one departs from the horizon. It is rather natural to expect that $A_{\ell m}^{\rm out} \to 0$  (or, to be more precise,
$A_{\ell m}^{\rm out}/B_{\ell m}^{\rm out} \to 0$)
for small  $\omega$, and that the near-horizon modes $B_{\ell m}^{\rm in}$ and $B_{\ell m}^{\rm out}$ could
be considered as long-living in the sense that the tunneling probability to infinity is extremely low, implying that near-horizon low-energy perturbations of the scalar fields   tend to be confined near the black hole horizon.  Moreover, since they are long-living and spatially confined, 
it is also natural to assume  
that such near-horizon modes could in principle attain thermal equilibrium, possibly with the black hole Hawking temperature $T_H=1/8\pi M$. 

Our analysis is based on the assumption that 
 the 
  dynamics of the near-horizon   $B_{\ell m}^{\rm in}$ and $B_{\ell m}^{\rm out}$  modes, for small $\omega$,
  can be well approximated by employing  the  Schr\"odinger operator
\begin{equation}
\label{Schr1}
{\mathcal{H}} = -\frac{d^2}{dr_*^2} + V^{\rm nh}_\ell(r_*)
\end{equation}
on the domain $\left(\left.-\infty , r_*^{\rm max}\right.\right]$, for some finite $r_*^{\rm max}$ corresponding to a $r$ not far from the horizon $r=2M$.  This is, of course, equivalent to assume that, for small $\omega$,
$A_{\ell m}^{\rm out} = 0$, leading to a perfect reflection due to the effective potential barrier and, consequently, to a confinement  of the near-horizon modes.
This approach closely resembles the so-called ``brick wall'' proposal for the thermodynamical analysis of fields around black holes \cite{BW}, even though we are concerned here with the dynamics in the interior region of the wall. As we will see,   our approach may indeed be considered 
a generalization of the standard brick wall hypothesis. 

It is a well known problem in standard Quantum Mechanics that the free-particle Schr\"odinger operator on the 
semiaxis has infinitely many self-adjoint extensions parameterized by a phase $\theta\in U(1)$, see \cite{AJP,GTV}, for instance, for further references.
We will show that similar results  also hold for our problem, {\em i.e.}, the Schr\"odinger operator (\ref{Schr1}) on the domain $\left(\left.-\infty , r_*^{\rm max}\right.\right]$  has infinitely many self-adjoint extensions determined by the boundary condition at $r_*^{\rm max}$.  Moreover, all  self-adjoint extensions in this case will give origin to physically acceptable descriptions for the near-horizon modes. Nevertheless, the  comparison with a simplified model suggests a physical prescription to
chose the pertinent  extensions.

\section{Self-adjoint extensions of the \\ effective Schr\"odinger operator}

Let us introduce the dimensionless variable $x = r_*/2M$, in terms of which one has the following
effective Schr\"odinger equation for near-horizon modes
\begin{equation}
\label{Schr2}
{\mathcal{H}} u_{\ell m} = \left( -\frac{d^2}{dx^2} + \frac{c_\ell^2}{4} e^{x} \right)u_{\ell m} =
\lambda^2 u_{\ell m},
\end{equation}
where
\begin{equation}
\label{cl}
c_\ell^2 = \frac{4}{e}\left(\ell^2+\ell+1\right)
\end{equation}
 and
$\lambda =2M\omega$, which we assume initially to be positive. (The possibility of having
imaginary $\lambda$     will be discussed in the last section. ) The 
  functions $u_{\ell m}$ are defined over the domain $\left(\left.-\infty , x^{\rm max}\right.\right]$.
  As we will see, our conclusions are independent of the precise value of $x^{\rm max}$, provided, of course, it is finite. We will drop the indices $\ell$ and $m$ for all functions and  constants   hereafter. 
It is natural to consider the initial domain $D({\mathcal{H}})$ of the effective Schr\"odinger operator (\ref{Schr2}) as   $C^\infty_0\left(\left.-\infty , x^{\rm max}\right.\right]$, {\em i.e.}, the smooth (complex) functions $u$ with compact support on the domain $\left(\left.-\infty , x^{\rm max}\right.\right]$. Notice that ${\mathcal{H}}$ is a symmetric operator with respect to the inner product 
\begin{equation}
\label{inner}
\left\langle v,u \right\rangle = \int_{-\infty}^{x^{\rm max}} \bar vu \, dx
\end{equation}
since 
\begin{equation}
\label{symm}
\left\langle v, {\mathcal{H}}u \right\rangle  = \left\langle  {\mathcal{H}}v,u \right\rangle 
\end{equation}
for all $u,v\in  D({\mathcal{H}})$. However, it is clear too that $D({\mathcal{H}}) \subset D({\mathcal{H}^\dagger})$ since (\ref{symm}) is valid  also for functions $ v\notin  D({\mathcal{H}})$, and this is the start point of the self-adjointness analysis of unbounded operators on Hilbert spaces \cite{AJP,GTV}.
On physical grounds, we should expect $D({\mathcal{H}^\dagger})$ to be the set of all smooth functions with finite norm $||v|| = \sqrt{\left\langle v,v \right\rangle}$, or at least finite norm per length unit in order to accommodate some possible plane wave solutions. Hence, we will consider  $D({\mathcal{H}^\dagger})$ as the set of smooth functions $v\in L^2\left(\left.-\infty , x^{\rm max}\right.\right]$, with the norm induced by (\ref{inner}).
 The von Neumann theorem assures that ${\mathcal{H}}$ will admit self-adjoint extensions provided the so-called deficiency index $n_+$ and $n_-$ be equal and greater than zero, 
where $n_\pm$ are the dimension of the deficiency subspaces $N_\pm \subset D({\mathcal{H}^\dagger})$ defined by
\begin{equation}
N_\pm = \left\{ v\in  D({\mathcal{H}^\dagger}), \quad {\mathcal{H}}v = \pm i v\right\}.
\end{equation}
In order to determine the 
   deficiency subspaces $N_\pm$, notice that the change of variable $z=e^\frac{x}{2}$ reduces (\ref{Schr2})   to a modified Bessel equation, allowing us to write down the general solution of  ${\mathcal{H}}v = \pm i v$ in terms of standard modified Bessel functions
\begin{equation}
 v(x) = a I_{\mu_\pm}\left( c e^\frac{x}{2} \right) + bK_{\mu_\pm}\left( c e^\frac{x}{2} \right),
\end{equation}
where $a$ and $b$ are constants and 
\begin{equation}
\mu_\pm = \sqrt{2}\left( 1 \mp i\right).
\end{equation}
From the standard asymptotic expressions for modified Bessel functions \cite{Abramo-Stegun}, one has for $x\to-\infty$
\begin{equation}
I_{\mu_\pm}\left( c e^\frac{x}{2} \right) \approx 
\frac{\left(\frac{c}{2}\right)^{\sqrt{2}\left( 1 \mp i\right)}}{\sqrt{2}\left( 1 \mp i\right)\Gamma\left(\sqrt{2}\left( 1 \mp i\right)\right) }e^{\frac{1\mp i}{\sqrt{2}}x}
\end{equation}
and
\begin{equation}
K_{\mu_\pm}\left( c e^\frac{x}{2} \right) \approx \frac{1}{2} 
 \left(\frac{c}{2}\right)^{\sqrt{2}\left( 1 \mp i\right)} \Gamma\left(\sqrt{2}\left( 1 \mp i\right)\right)e^{-\frac{1\mp i}{\sqrt{2}}x}.
\end{equation}
It is clear that the modified Bessel function $K_{\mu_\pm}$ will give origin to solutions $v \notin  D({\mathcal{H}^\dagger})$ since they will diverge exponentially for $x\to-\infty$. Hence, only the solutions involving $I_{\mu_\pm}$ are allowed, and we have $n_+ =n_- = 1$. The deficiency subspaces $N_\pm$ are then vector spaces with dimension 1 generated by $I_{\mu_\pm}$, and  von Neumann theorem assures that ${\mathcal{H}}$ has a family of
self-adjoint extensions parameterized by a phase $\theta \in U(1)$  \cite{AJP,GTV}.

The structure of the differential operator ${\mathcal{H}}$ 
is rather simple and   will allow us to determine explicitly all of its self-adjoint extensions ${\mathcal{H}}_\alpha$. Notice that, for smooth $u,v\in L^2\left(\left.-\infty , x^{\rm max}\right.\right]$, one has
\begin{equation}
\left\langle v, {\mathcal{H}}u \right\rangle  - \left\langle  {\mathcal{H}}v,u \right\rangle =
\overline{v'}(x^{\rm max})u(x^{\rm max}) - \overline{v}(x^{\rm max}){u'}(x^{\rm max}),
\end{equation}
from where we see that ${\mathcal{H}}$  will be self-adjoint provided 
\begin{equation}
\frac{v(x^{\rm max})}{v'(x^{\rm max})}=\frac{u(x^{\rm max})}{u'(x^{\rm max})} = \alpha = \tan\frac{\theta}{2},
\end{equation}
with $\theta \in (-\pi,\pi)$, and we have finally established 
\begin{equation}
D\left({\mathcal{H}}_\alpha\right) = D\left({\mathcal{H}}_\alpha^\dagger\right) =\left\{ v\in   L^2\left(\left.-\infty , x^{\rm max}\right.\right] \left| \quad v(x^{\rm max}) = \alpha v'(x^{\rm max})\right.\right\},
\end{equation}
with $ {\mathcal{H}}_\infty  $   corresponding to the boundary condition
$v'(x^{\rm max})=0$. It is worthy to notice that the case $ {\mathcal{H}}_0  $, which corresponds to 
$v(x^{\rm max})=0$,   corresponds to the  brick wall  hypothesis \cite{BW}. Our analysis, besides of involving more general boundary conditions, is  restricted to the other side of the wall, {\em i.e.} to the modes confined in the near-horizon region. Notice that the differential expression for the operator ${\mathcal{H}}_\alpha$ is independent of $\alpha$, it alters only $D({\mathcal{H}})$.

 In order to interpret the physical meaning of the self-adjoint extensions $ {\mathcal{H}}_\alpha $ , let us consider now 
the  eigenproblem (\ref{Schr2}) for positive $\lambda$. It has also solutions in terms of modified Bessel functions $I_\mu$ and $K_\mu$, but now with pure imaginary order $\mu = 2i\lambda$. However, it is more convenient for our purposes here to write down the solution as a linear combination of $I_{2i\lambda}$ and $\overline{I_{2i\lambda}} = I_{-2i\lambda}$. One has
\begin{equation}
\label{sol1}
u(x) = a_\lambda I_{2i\lambda}\left( c e^\frac{x}{2} \right) + b_\lambda I_{-2i\lambda}\left( c e^\frac{x}{2} \right),
\end{equation}
with $a_\lambda $ and $b_\lambda $ constants. 
For $x\to-\infty$, we have \cite{Abramo-Stegun}
\begin{equation}
u(x) \approx \frac{a_\lambda\left(\frac{c}{2}\right)^{2i\lambda} }{2i\lambda\Gamma(2i\lambda)}e^{i\lambda x} -
 \frac{b_\lambda\left(\frac{c}{2}\right)^{-2i\lambda} }{2i\lambda\Gamma(-2i\lambda)}e^{-i\lambda x},
\end{equation} 
from where one can read the scattering coefficients in the region very close to the horizon
\begin{equation}
\label{coeff}
B^{\rm in}_\lambda = -
 \frac{b_\lambda\left(\frac{c}{2}\right)^{-2i\lambda} }{2i\lambda\Gamma(-2i\lambda)},\quad {\rm and} \quad
 B^{\rm out}_\lambda = \frac{a_\lambda\left(\frac{c}{2}\right)^{2i\lambda}  }{2i\lambda\Gamma(2i\lambda)} .
\end{equation}
Defining the reflection coefficient as 
\begin{equation}
R_\lambda = \frac{B^{\rm in}_\lambda}{B^{\rm out}_\lambda} = -\frac{b_\lambda}{a_\lambda} \frac{\Gamma(2i\lambda)}{\Gamma(-2i\lambda)}\left(\frac{c}{2}\right)^{-4i\lambda},
\end{equation}
we have 
\begin{equation}
|R_\lambda| = \left|\frac{b_\lambda}{a_\lambda}\right|.
\end{equation}
On the other hand, one can determine $b_\lambda/a_\lambda$ from the boundary condition $u(x^{\rm max}) = \alpha u'(x^{\rm max})$. One has
\begin{equation}
\label{ratio}
\frac{b_\lambda}{a_\lambda} = -\frac{\chi}{\overline{\chi}}
\end{equation}
where
\begin{equation}
\label{chi}
\chi = I_{2i\lambda}\left( c e^{\frac{1}{2}x^{\rm max}} \right) -\frac{c\alpha}{2}e^{\frac{1}{2}x^{\rm max}}  I'_{2i\lambda}\left( c e^{\frac{1}{2}x^{\rm max}}\right),
\end{equation}
which clearly implies  that $|R_\lambda|=1$, meaning that, irrespective of the value of $\alpha$, we have always full reflection of the near-horizon modes on the effective potential barrier, which is compatible with  $A^{\rm out}_\lambda = 0$ as expected. 
From the scattering point of view, it is possible to implement a brick wall which effectively confine the modes in the near-horizon region without imposing $u(x^{\rm max})=0$. Moreover, any value of $\alpha$ is perfectly admissible in this context, all self-adjoint extensions give origin to physically acceptable descriptions for the near-horizon modes.
 We will have a complete set of (continuous) eigenvalues and eigenvectors for (\ref{Schr2}) for any value of $\alpha$. As we will see below, all self-adjoint extensions will lead also to consistent thermodynamics for the near-horizon modes. 

\subsection{Statistical mechanics and thermal equilibrium }

All self-adjoint extensions ${\mathcal{H}}_\alpha$ describe confined incoming and outcoming near-horizon modes characterized by
the coefficients $B^{\rm in}_\lambda$ and $B^{\rm out}_\lambda $, see (\ref{coeff}). The probability of having incoming and outcoming modes with energy $\lambda  $ in the horizon are, respectively,
\begin{equation}
\big|B^{\rm in}_\lambda  \big|^2 =   {\frac{ \sinh 2\pi\lambda}{2\pi\lambda}} |b_\lambda|^2 ,\quad
\left|B^{\rm out}_\lambda \right|^2 =   {\frac{ \sinh 2\pi\lambda}{2\pi\lambda}}|a_\lambda|^2 . 
\label{prob}
\end{equation}
Notice that for small $\lambda$ we have essentially $\big|B^{\rm in}_\lambda  \big|^2 \approx|b_\lambda|^2$ and
$\big|B^{\rm out}_\lambda  \big|^2 \approx |a_\lambda|^2$. Let us suppose now that the near-horizon modes are at thermal equilibrium with temperature $T=\tau/2M$ (the Hawking temperature corresponds to $\tau = 1/4\pi)$. Assuming a grand canonical ensemble and the detailed balance principle \cite{DB}, we expect that incoming and outcoming modes be separately at thermal equilibrium, meaning that we should expect
 that both $ |B^{\rm in}_\lambda  \big|^2 $ and $ |B^{\rm out}_\lambda  \big|^2 $ obey Boltzmann distributions and, hence, both should be proportional to $e^{-\lambda/\tau}$. Interestingly, such detailed balance condition, which implies that incoming and outcoming modes are equally probable in a regime of thermal equilibrium, is compatible with any value of $\alpha$, {\rm i.e.}, all self-adjoint extensions
  ${\mathcal{H}}_\alpha$  are equivalent also from the thermal equilibrium point of view. The compatibility is assured by the fact that $|a_\lambda|^2= |b_\lambda|^2$ for any value of $\alpha$, see (\ref{ratio}) and (\ref{chi}). Hence, if one of the modes is assumed to be at thermal equilibrium, by (\ref{prob}) the other automatically be also at thermal equilibrium. It is fundamental for the detailed balance that the 
  boundary condition implies 
\begin{equation}
b_\lambda =  e^{i\psi_\lambda }a_\lambda,
\end{equation}
where the phase $\psi_\lambda$ depends on all   parameters of the problem, 
see  (\ref{ratio}) and (\ref{chi}), and particularly on the energy  $\lambda$. Nevertheless, irrespective of the value of $\alpha$, we have always $|a_\lambda|^2= |b_\lambda|^2$.

\subsection{A prescription for the extension selection}
Rigorously, for each value of $\alpha$ we have a fixed domain on the Hilbert space and a complete, physically consistent, description for the 
low-energy modes. We should not mix modes with different $\alpha$ since they belong to different domains. The physical interpretation of the parameter $\alpha$  is still rather unclear, but a simplified model can help to shed some light here. Let us consider the well-known elementary problem of the scattering by a rectangular barrier
\begin{equation}
\label{rect}
V(x) = \left\{
\begin{array}{cc} 
0, & x< 0,\\
V_0, & 0\le x \le L, \\
0, & x > L,
\end{array}
 \right.
\end{equation}
with both $V_0$ and $L$ positives. We are interested on  scattering problems of the type (\ref{cond}), {\em i.e.}, on solutions of the type
\begin{equation}
\label{sol2}
u(x) = \left\{ 
\begin{array}{cc}\displaystyle
B^{\rm in}_{\lambda} e^{-i\lambda x} +
B^{\rm out}_{\lambda} e^{i\lambda x} , & x < 0 , \\
C_\lambda e^{\sqrt{V_0 - \lambda^2}x} + D_\lambda e^{-\sqrt{V_0 - \lambda^2}x} , & 0\le  x \le L \\
A^{\rm out}_\lambda e^{i\lambda x}
, & x> L ,
\end{array}
\right.
\end{equation}
with $\lambda^2 < V_0$. The standard matching conditions at $x=0$ and $x=L$ read
\begin{eqnarray}
B^{\rm in} + B^{\rm out} &=& C_\lambda + D_\lambda,\\ 
-i\lambda \left(B^{\rm in} - B^{\rm out}\right) &=& \sqrt{V_0-\lambda^2} ( C_\lambda - D_\lambda),\\
 A^{\rm out}_\lambda e^{i\lambda L}  &=&  C_\lambda e^{\sqrt{V_0 - \lambda^2}L} + D_\lambda e^{-\sqrt{V_0 - \lambda^2}L},\\
 i\lambda  A^{\rm out}_\lambda e^{i\lambda L} &=& \sqrt{V_0-\lambda^2} \left( C_\lambda e^{\sqrt{V_0 - \lambda^2}L} - D_\lambda e^{-\sqrt{V_0 - \lambda^2}L}\right).
\end{eqnarray}
After some straightforward algebra, one can evaluate the usual reflection coefficient $R_\lambda $ leading to
\begin{equation}
\left| R_\lambda \right|^2 =   \frac{V_0\sinh^2 \sqrt{V_0-\lambda^2} L }   { 4\lambda^2 (V_0-\lambda^2) +  V_0\sinh^2 \sqrt{V_0-\lambda^2} L }   .
\end{equation}
The problem of near-horizon modes is mimicked in this toy model by assuming $L\to\infty$, which implies $|R_\lambda|\to 1$, {\em i.e.}, full reflection leading to a ``confinement'' of the solutions (\ref{sol2}) in the negative semiaxis. Since $|R_\lambda|\to 1$, we know that $A^{\rm out}_\lambda \to 0$ and hence from (35) and (36) we have that $C_\lambda \to 0$, which implies the following condition for $u(x)$ on $x=0$
\begin{equation}
u'(0) = -\sqrt{V_0 - \lambda^2}u(0).
\end{equation} 
Thus, finally, in the low-energy limit, $\lambda^2 \ll V_0$, we have that the dynamics of the totally reflect solutions for the barrier (\ref{rect}) may be viewed as an effective Schr\"odinger equation for a free particle on the negative semiaxis with the boundary condition corresponding to $\alpha^{-1} = -\sqrt{V_0}$. This simple results suggests that $\alpha^{-1} = -\sqrt{\max V_\ell} $ for the near-horizon modes. We would have different self-adjoint extensions for different angular momentum numbers $\ell$, but this is hardly a surprise since the effective potential (\ref{pot}), and consequently the Schr\"odinger operator (\ref{Schr2}), does depend explicitly on $\ell$. 
It is interesting to notice that the standard brick wall condition $\alpha = 0$  would require $\max V_\ell\to\infty$, which on the other hand demands $\ell\to \infty$. Nevertheless, all self-adjoint extensions act effectively as brick walls since we have full reflection for all values of $\alpha$.  In fact, despite
our prescription for the selection of $\alpha$, since all extensions are in
principle physically equivalent, they might be considered in equal footing for 
statistical analyses of near-horizon modes around black holes

\section{Final remarks} 

We will revisit in this last section two  previously noticed points. First, that our results do not 
depend on the details of the Schwarzchild black hole. They will also hold for any non-extremal,
 spherically symmetric, static, and asymptotically flat  black hole. 
The metric of a generic  spherically symmetric static black hole  can be cast in the form
\begin{equation}
\label{metric}
ds^2 = -f(r)dt^2 + \frac{dr^2}{h(r)} + r^2d\Omega^2.
\end{equation}
The event horizon corresponds to the outermost zero of $f(r)$, say at $r=r_0$. The black hole is said to be non-extremal if $f'(r_0) = k >  0$, and hence in the vicinity of the horizon we have $f(r) \approx k_1(r-r_0)$. Regularity of the horizon area demands a smooth $\sqrt{-g}$, and from (\ref{metric}) we see   also that $h(r) \approx k_2(r-r_0)$, with $k_2>0$. By using the standard decomposition (\ref{decomposition}) for the Klein-Gordon equation on the metric (\ref{metric}), we arrive to a Scḧ́\"odinger-like equation as (\ref{Schr1}), but now with the effective potential
\begin{equation}
\label{newpot}
\tilde V_\ell(r ) = \ell(\ell+1)\frac{f}{r^2} + \frac{1}{2r}\left(f'h + fh' \right)
\end{equation}
and tortoise coordinates such that
\begin{equation}
\label{tort}
\frac{d r_*}{dr} = \frac{1}{\sqrt{fh}}.
\end{equation}
If (\ref{metric}) is assumed to be asymptotically flat, we have   $f(r)\to 1$ and $h(r)\to 1$   for $r\to \infty $, and hence (\ref{newpot}) decays as a power law at infinity in the same way the Schwarzschild potential (\ref{pot}) does. On the other hand, in the near-horizon region  one has
\begin{equation}
\tilde V_\ell(r ) \approx k_1(r-r_0)\left( \frac{\ell(\ell +1) }{r_0^2} + \frac{  k_2}{r_0} \right). 
\end{equation}
The new tortoise coordinate (\ref{tort}) also obeys $r_*\to -\infty$ on the horizon and, moreover, we have
\begin{equation}
r-r_0 = r_0 e^{\sqrt{k_1k_2}r_*},
\end{equation}
from where we conclude that the effective potential (\ref{newpot}) also decays exponentially in the near-horizon region. Indeed, the aspect of the generic effective potential (\ref{newpot})
of a non-extremal, spherically symmetric, static, and asymptotically flat black hole 
 is qualitatively the same of the Schwarzschild case, Fig. \ref{fig1}. All the analyses we have done
  follow 
 analogously for the generic black hole  case. 

The second point corresponds to the imaginary $\lambda$ case in (\ref{Schr2}). It is a well known and curious fact that the Schr\"odinger equation for the free particle on the semiaxis admits some bounded solutions, with negative energy, for certain self-adjoint extension choices, see \cite{AJP}. We have the same interesting behavior here and they indeed correspond to the imaginary $\lambda$ solutions of the eigenproblem (\ref{Schr2}). For $\lambda = \sigma i$, the fundamental solutions of (\ref{Schr2}) will be linear combinations of the modified Bessel functions $I_{2\sigma} $ and $K_{2\sigma}$.  From the asymptotic behavior near the origin, we can discharge the second solution. Using the standard series expansion \cite{Abramo-Stegun} for  $I_{2\sigma} $, we have the following solution for the eigenproblem (\ref{Schr2}) with eigenvalue $\lambda^2 = -\sigma^2$,
\begin{equation}
\label{series}
u(x) = a_\sigma I_{2\sigma}\left(ce^\frac{x}{2} \right) =  \sum_{k=0}^\infty \frac{e^{(k+\sigma)x}}{k!\Gamma(k+2\sigma+1)}\left(\frac{c}{2}\right)^{2(k+\sigma)} ,
\end{equation}
where it is assumed $\sigma > 0$. 
It is clear from (\ref{series}) that $u(x)$ and all of its derivative are monotonically increasing functions and, thus, in order to accommodate such bounded solution for (\ref{Schr2}), a self-adjoint extension with $\alpha > 0$ is required, which will never be selected by our prescription. In our case, such bounded solutions do not oscillate, see (\ref{decomposition}), but rather decrease  exponentially. This kind of overdamped evolution for scalar fields is quite similar to some highly
damped quasinormal modes that are known to exist for generic black holes, see \cite{Damped}. This topic is now under investigation.

\section*{Acknowledgements}
ARQ and AS thank the University of Zaragoza, where part of this work was carried on, for the warm hospitality. The authors acknowledge the financial support of CNPq and CAPES (ARQ and AS) and FAPESP (AS, Grant 2013/09357-9).

\end{document}

%% file: plot.tex
\begingroup
  \makeatletter
  \providecommand\color[2][]{%
    \GenericError{(gnuplot) \space\space\space\@spaces}{%
      Package color not loaded in conjunction with
      terminal option `colourtext'%
    }{See the gnuplot documentation for explanation.%
    }{Either use 'blacktext' in gnuplot or load the package
      color.sty in LaTeX.}%
    \renewcommand\color[2][]{}%
  }%
  \providecommand\includegraphics[2][]{%
    \GenericError{(gnuplot) \space\space\space\@spaces}{%
      Package graphicx or graphics not loaded%
    }{See the gnuplot documentation for explanation.%
    }{The gnuplot epslatex terminal needs graphicx.sty or graphics.sty.}%
    \renewcommand\includegraphics[2][]{}%
  }%
  \providecommand\rotatebox[2]{#2}%
  \@ifundefined{ifGPcolor}{%
    \newif\ifGPcolor
    \GPcolortrue
  }{}%
  \@ifundefined{ifGPblacktext}{%
    \newif\ifGPblacktext
    \GPblacktexttrue
  }{}%
  \let\gplgaddtomacro\g@addto@macro
  \gdef\gplbacktext{}%
  \gdef\gplfronttext{}%
  \makeatother
  \ifGPblacktext
    \def\colorrgb#1{}%
    \def\colorgray#1{}%
  \else
    \ifGPcolor
      \def\colorrgb#1{\color[rgb]{#1}}%
      \def\colorgray#1{\color[gray]{#1}}%
      \expandafter\def\csname LTw\endcsname{\color{white}}%
      \expandafter\def\csname LTb\endcsname{\color{black}}%
      \expandafter\def\csname LTa\endcsname{\color{black}}%
      \expandafter\def\csname LT0\endcsname{\color[rgb]{1,0,0}}%
      \expandafter\def\csname LT1\endcsname{\color[rgb]{0,1,0}}%
      \expandafter\def\csname LT2\endcsname{\color[rgb]{0,0,1}}%
      \expandafter\def\csname LT3\endcsname{\color[rgb]{1,0,1}}%
      \expandafter\def\csname LT4\endcsname{\color[rgb]{0,1,1}}%
      \expandafter\def\csname LT5\endcsname{\color[rgb]{1,1,0}}%
      \expandafter\def\csname LT6\endcsname{\color[rgb]{0,0,0}}%
      \expandafter\def\csname LT7\endcsname{\color[rgb]{1,0.3,0}}%
      \expandafter\def\csname LT8\endcsname{\color[rgb]{0.5,0.5,0.5}}%
    \else
      \def\colorrgb#1{\color{black}}%
      \def\colorgray#1{\color[gray]{#1}}%
      \expandafter\def\csname LTw\endcsname{\color{white}}%
      \expandafter\def\csname LTb\endcsname{\color{black}}%
      \expandafter\def\csname LTa\endcsname{\color{black}}%
      \expandafter\def\csname LT0\endcsname{\color{black}}%
      \expandafter\def\csname LT1\endcsname{\color{black}}%
      \expandafter\def\csname LT2\endcsname{\color{black}}%
      \expandafter\def\csname LT3\endcsname{\color{black}}%
      \expandafter\def\csname LT4\endcsname{\color{black}}%
      \expandafter\def\csname LT5\endcsname{\color{black}}%
      \expandafter\def\csname LT6\endcsname{\color{black}}%
      \expandafter\def\csname LT7\endcsname{\color{black}}%
      \expandafter\def\csname LT8\endcsname{\color{black}}%
    \fi
  \fi
    \setlength{\unitlength}{0.0500bp}%
    \ifx\gptboxheight\undefined%
      \newlength{\gptboxheight}%
      \newlength{\gptboxwidth}%
      \newsavebox{\gptboxtext}%
    \fi%
    \setlength{\fboxrule}{0.5pt}%
    \setlength{\fboxsep}{1pt}%
\begin{picture}(6480.00,4536.00)%
    \gplgaddtomacro\gplbacktext{%
      \csname LTb\endcsname
      \put(814,704){\makebox(0,0)[r]{\strut{}$0$}}%
      \put(814,1065){\makebox(0,0)[r]{\strut{}$0.2$}}%
      \put(814,1426){\makebox(0,0)[r]{\strut{}$0.4$}}%
      \put(814,1787){\makebox(0,0)[r]{\strut{}$0.6$}}%
      \put(814,2148){\makebox(0,0)[r]{\strut{}$0.8$}}%
      \put(814,2510){\makebox(0,0)[r]{\strut{}$1$}}%
      \put(814,2871){\makebox(0,0)[r]{\strut{}$1.2$}}%
      \put(814,3232){\makebox(0,0)[r]{\strut{}$1.4$}}%
      \put(814,3593){\makebox(0,0)[r]{\strut{}$1.6$}}%
      \put(814,3954){\makebox(0,0)[r]{\strut{}$1.8$}}%
      \put(814,4315){\makebox(0,0)[r]{\strut{}$2$}}%
      \put(946,484){\makebox(0,0){\strut{}$-6$}}%
      \put(1680,484){\makebox(0,0){\strut{}$-4$}}%
      \put(2414,484){\makebox(0,0){\strut{}$-2$}}%
      \put(3148,484){\makebox(0,0){\strut{}$0$}}%
      \put(3881,484){\makebox(0,0){\strut{}$2$}}%
      \put(4615,484){\makebox(0,0){\strut{}$4$}}%
      \put(5349,484){\makebox(0,0){\strut{}$6$}}%
      \put(6083,484){\makebox(0,0){\strut{}$8$}}%
    }%
    \gplgaddtomacro\gplfronttext{%
      \csname LTb\endcsname
      \put(198,2509){\rotatebox{-270}{\makebox(0,0){\strut{}Rescaled effective potential $4M^2V_\ell\left(\frac{r_*}{2M}\right)$}}}%
      \put(3514,154){\makebox(0,0){\strut{}Rescaled tortoise coordinate $\frac{r_*}{2M}$}}%
      \csname LTb\endcsname
      \put(5096,2729){\makebox(0,0)[r]{\strut{}$\ell=1$}}%
      \csname LTb\endcsname
      \put(5096,2509){\makebox(0,0)[r]{\strut{}$\ell=2$}}%
      \csname LTb\endcsname
      \put(5096,2289){\makebox(0,0)[r]{\strut{}$\ell=3$}}%
    }%
    \gplbacktext
    \put(0,0){\includegraphics{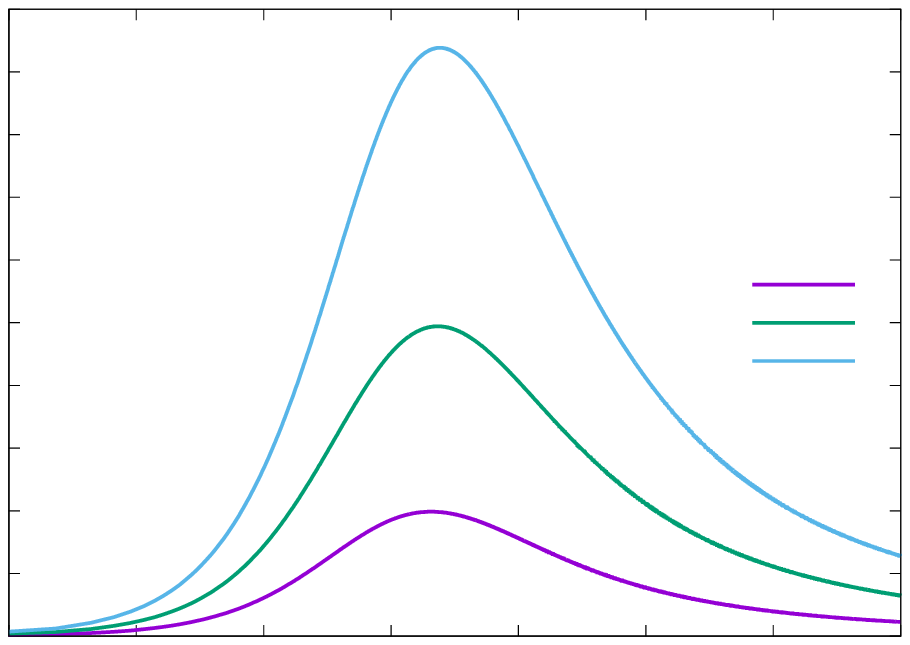}}%
    \gplfronttext
  \end{picture}%
\endgroup